\documentclass[
    ,final            
  ]
  {aipproc}

\layoutstyle{6x9}

\begin{document}

\title[Dynamics of faint stellar populations]{Studying the kinematics of faint stellar populations with the Planetary Nebula Spectrograph}

\classification{98.10.+z, 98.62.Dm}
\keywords      {Galaxy dynamics, Planetary nebulae}

\author{Michael R. Merrifield}{
  address={School of Physics \& Astronomy, University of Nottingham,
    UK},
  email={michael.merrifield@Nottingham.ac.uk}
}

\author{The PN.S Consortium}{
  address={\url{http://www.astro.rug.nl/~pns/}}
}

\begin{abstract}
  Galaxies are faint enough when one observes just their light
  distributions, but in studying their full dynamical structure the
  stars are spread over the six dimensions of phase space rather than
  just the three spatial dimensions, making their densities very low
  indeed.  This low signal is unfortunate, as stellar dynamics hold
  important clues to these systems' life histories, and the issue is
  compounded by the fact that the most interesting information comes
  from the faintest outer parts of galaxies, where dynamical
  timescales (and hence memories of past history) are longest.

  To extract this information, we have constructed a special-purpose
  instrument, the Planetary Nebula Spectrograph, which observes
  planetary nebulae as kinematic tracers of the stellar population,
  and allows one to study the stellar dynamics of galaxies down to
  extremely low surface brightnesses.  Here, we present results from
  this instrument that illustrate how it can uncover the nature of low
  surface-brightness features such as thick disks by studying their
  kinematics, and trace faint kinematic populations that are
  photometrically undetectable.
\end{abstract}

\maketitle


\section{Introduction}

Because stars are well-spaced, even the inner parts of galaxies can
appear faint, comparable in surface brightness to the darkest of night
skies.  Not surprisingly, then, when we start exploring the full
six-dimensional phase space of velocities and positions that stars
occupy in galaxies, the signal per phase space element becomes almost
vanishingly small.  In the case of the Milky Way, for example, the
fascinating stellar streams in the halo \citep{Belokurovetal06} have
widths of only $\sim 20\,{\rm pc}$, and velocity dispersions of maybe
$\sim 10\,{\rm km}\,{\rm s}^{-1}$.  If we were to divide the Galaxy,
with a linear dimension of $\sim 20\,{\rm kpc}$ and velocities
spanning $\sim 600\,{\rm km}\,{\rm s}^{-1}$, into resolution elements
this small, we would end up with $\sim 10^{14}$ of them, many more
than there are stars in the Milky Way, so the average occupation
number per element would be very small indeed.

Thus, if we are to study the detailed kinematics of galaxies, we need
to do two things.  First, we must be able to detect stellar
populations at extremely low phase densities, which essentially means
detecting individual stars.  Second, we need to come up with ways to
combine the information from these individual detections in order to
bring together enough signal to say anything about the properties of
the galaxy, and hence reconstruct its detailed dynamical structure in
order to learn about its evolution.

Planetary nebulae (PNe) offer an ideal tracer of such faint stellar
populations, as they are readily individually detectable from their
emission lines, even in quite distant galaxies, and the Doppler shifts
in these lines can be used to measure their line-of-sight velocities.
To make such measurements efficiently in a single step, we have
designed and built a customized instrument, the Planetary Nebula
Spectrograph or PN.S \citep{Douglasetal02}.  In this paper, we
present examples from a couple of the projects that we have been
undertaking with this instrument, to illustrates the principles of
investigating galaxies in the ultimate low-density regime of phase
space.

\section{The Thick Disk in NGC~891}

When one traces edge-on stellar disks away from the plane out to very
low surface brightnesses, an excess of light above the extrapolation
of the normal disk population is frequently found
\citep{YoachimDalcanton06}.  The nature of such ``thick disks''
remains highly controversial: they could just be the extreme tail of
the normal disk population, or they could represent some more dramatic
event, such as the debris from a merger.  One exciting new clue to
help distinguish between these possibilities was uncovered when it was
found that the thick disk in the edge-on system FGC~227 appears
to be rotating in the opposite direction to the normal thin-disk
population, indicating that it would have to be a distinct component
such as might be formed in a merger \citep{YoachimDalcanton05}.
However, the surface brightness of this component is so faint that,
even with ten hours of integration on an 8-metre telescope,
conventional spectroscopy had such large error bars that the data were
also consistent with co-rotating thin and thick disks.

\begin{figure}
  \includegraphics[width=9.25cm]{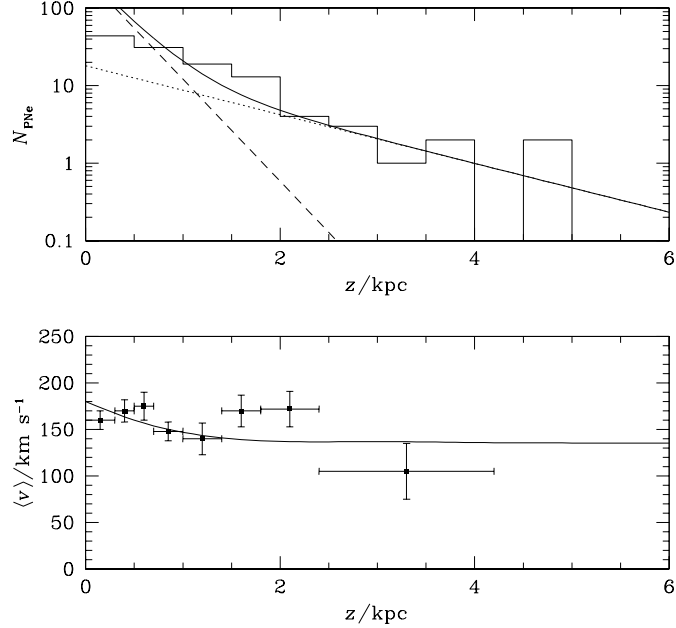}
  \label{fig:NGC891}
  \caption{Properties of the PNe in the disk of the edge-on galaxy
    NGC~891 as a function of distance from the galactic plane.  The
    upper panel shows the number counts of PNe as detected with PN.S,
    with lines showing a model decomposition of conventional
    photometry into thin disk and thick disk \citep{Morrisonetal97}.
    The lower panel shows the mean rotation velocity of these PNe,
    with a line showing the simplest possible single-population model
    for their kinematics.}
\end{figure}

As Figure~\ref{fig:NGC891} illustrates, we can do a great deal better
with a rather shorter integration on a 4-metre telescope using PN.S.
The upper panel shows the number counts of PNe detected in the edge-on
spiral NGC~891 as a function of distance from its plane.  The lines
show a possible decomposition of the disk into thin and thick
exponential components \citep{Morrisonetal97}.  The data drop below
the model close to the plane, as PNe are difficult to detect against
the high surface brightness of the galaxy in this region, and some PNe
will also be lost in NGC~891's strong dust lane.  However, in the
region we are interested in away from the plane, the PN number counts
trace the light distribution very well, underlining their nature as
generic tracers of the stellar population.  Further, we are clearly
detecting them in significant numbers out to beyond the transition
where the thick disk becomes the dominant component.

The lower panel in this figure shows the observed mean rotation speed
for these PNe as a function of distance from the plane.  There is
plenty of signal to distinguish between co- and counter-rotating thin
and thick disk components, and in this system the two are clearly
rotating in the same direction.  Indeed, we can go further and fit a
simple single-component kinematic model to the data.  This heuristic
model is of the form
\begin{equation}
\langle v \rangle = \alpha(v_c - \beta \sigma_z(z)^2/v_c),
\end{equation}
where $\alpha$ provides the factor that allows for the line-of-sight
projection effects that reduce the line-of-sight velocity below the
mean rotational velocity \citep{Neisteinetal99}, while $\beta$
parameterizes all the terms related to the shape of the velocity
ellipsoid in the asymmetric drift equation \citep{BinneyTremaine08};
here it is assumed to be a constant.  The circular speed of the
galaxy, $v_c$, can be inferred from gas kinematics, while the vertical
velocity dispersion, $\sigma_z(z)$, is estimated by solving the Jeans
equation for a self-gravitating sheet whose density distribution is
given by the double exponential model in the upper panel of
Figure~\ref{fig:NGC891}.  Clearly, the data are essentially consistent
with this simple single-component model: there is no kinematic
evidence that the thick disk in NGC~891 is in any way a distinct entity.

\begin{figure}
  \includegraphics[width=9.25cm]{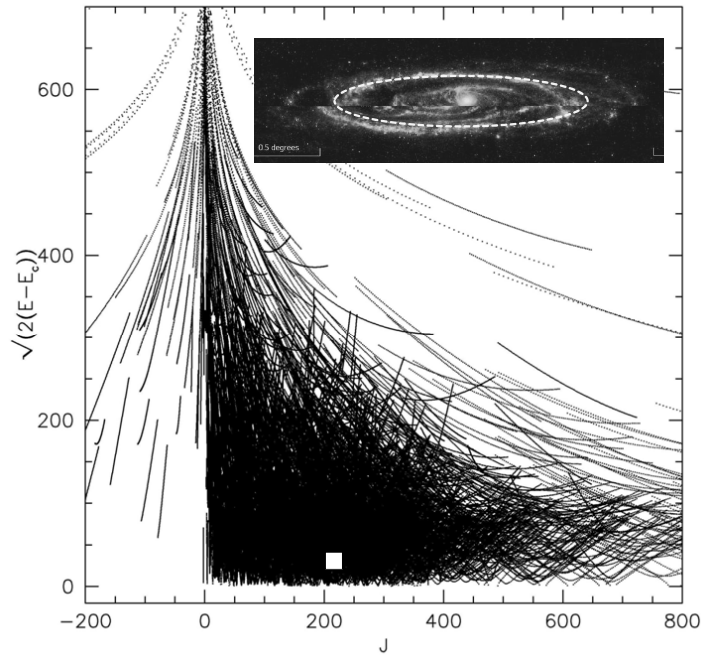}
  \caption{Plot of the possible values of angular momentum and energy
    (with the local circular energy subtracted, so a circular orbit
    lies at the bottom of the plot) for each of the PNe in the M31
    survey data.  The white box shows the point of the largest local
    excess of PNe.  The inset shows the orbit to which this point
    corresponds, superimposed on a mid-infrared image of the galaxy
    \citep{Gordonetal04}.}
   \label{fig:M31}
\end{figure}

\section{A Faint Kinematic Population in M31}
As a further illustration of the way in which faint kinematic
components can be detected, Figure~\ref{fig:M31} shows the results of
an analysis of the 2615 PNe that were found using PN.S in a survey of
M31 \citep{Merrettetal06}.  In an inclined disk system like this, the
phase space is essentially reduced to four dimensions, two spatial and
two velocity.  We can measure three of these components, two spatial
and the line-of-sight velocity, so each PN has only one unknown
coordinate.  Thus, while one cannot measure the energy and angular
momentum of any single PN, there is only a one-dimensional family of
possibilities, as illustrated by the line sections in the figure.  If
group of PNe share a common orbit (and hence energy and angular
momentum), then their lines will all intersect at a single point,
creating a local excess in this plane.  Such an excess of crossing
lines is, indeed, found in this figure, at the point indicated by the
box.  The location of the orbit to which it corresponds does not
appear in any way special in an optical image of M31, but, as the
inset image in the figure shows, mid-infrared data reveal a bright
ring of emission at this radius, indicating that it lies at one of the
major orbital resonances.  Thus, either we have detected the
population of stars born in this ring of star formation, or, more
likely since PNe come from old stars, we are picking out the
relatively small population of objects that have become dynamically
trapped at this resonance.  Whatever their origin, these data again
illustrate the power of PN.S to identify kinematic components that are
at far too low a density in phase space to be detectable by more
conventional means.





\bibliographystyle{aipproc}   

\end{document}